\documentclass[12pt,psfig]{article}
\usepackage[centertags]{amsmath}
\usepackage{amssymb}
\newcommand{\Z}{{\mathbb Z}}
\begin{document}
\begin{flushright}
\baselineskip=12pt
{ACT-01-05}\\
{MIFP-05-02}\\
\end{flushright}
\def\IZ{Z\kern-.5em Z}

\begin{center}
{\LARGE \bf A Supersymmetric Flipped SU(5) Intersecting Brane World \\}
\vglue 1.00cm
{C.-M. Chen$^{\spadesuit}$ \footnote
{cchen@physics.tamu.edu}, G. V. Kraniotis$^{\spadesuit}$ \footnote
 {kraniotis@physics.tamu.edu} , V. E. Mayes$^{\spadesuit}$ \footnote{eric@physics.tamu.edu},\\ D. V. Nanopoulos $^{\spadesuit,\clubsuit,\diamondsuit}$\footnote{dimitri@physics.tamu.edu}, and J. W. Walker $^{\spadesuit}$ \footnote{jwalker@physics.tamu.edu}\\}
\vglue 0.2cm
    {$\spadesuit$ \it  George P. and Cynthia W. Mitchell Institute for
Fundamental Physics,\\
 Texas A$\&$M University, College Station TX,
77843, USA  \\}
{$\clubsuit$ \it Astroparticle Physics Group, Houston Advanced Research Center (HARC)\\
Mitchell Campus, Woodlands, TX 77381, USA\\}
{$\diamondsuit$ \it  Academy of Athens, Division of Natural Sciences,  \\}
{\it 28 Panepistimiou Avenue, Athens 10679, Greece }
\baselineskip=12pt

\vglue 2.5cm
ABSTRACT
\end{center}

We construct an $N=1$ supersymmetric three-family flipped $SU(5)$ model from
type IIA orientifolds on $ T^6/(\Z_2\times \Z_2)$ with D6-branes
intersecting at general angles. The spectrum contains a complete grand unified
and electroweak Higgs sector. In addition, it contains extra exotic matter
both in bi-fundamental and vector-like representations as well as two copies of
matter in the symmetric representation of $SU(5)$.

{\rightskip=3pc
\leftskip=3pc
\noindent
\baselineskip=20pt

}

\vfill\eject
\setcounter{page}{1}
\pagestyle{plain}
\baselineskip=14pt

\section{Introduction}

The intersecting D-brane world approach \cite{BDL,BACHAS,BERLIN,AFIRU1}
plays a prominent role in the attempts
of string phenomenologists to reproduce the standard model physics in a
convincing way from type II string theory.

A number of  consistent non-supersymmetric three-generation
standard-like models have been constructed in \cite{CIM,BKL1} (for
a complete set of references the reader should consult the
excellent reviews \cite{REVIEWS1,
REVIEWS2,REVIEWS3,REVIEWS4,DIETER}). Open strings that begin and
end on a stack of $M$ D-branes generate the gauge bosons of the
group $U(M)$ living in the world volume of the D-branes.
 So the
standard approach is to start with one stack of 3 D-branes,
another of 2, and $n$ other stacks each having just 1 D-brane,
thereby generating the gauge group $U(3) \times U(2) \times
U(1)^n$. The D4-, D5- or D6-branes wrap the three large spatial
dimensions and respectively 1-, 2- or 3-cycles of the
six-dimensional internal space (typically a torus $T^6$ or a
Calabi-Yau 3-fold). Fermions in bi-fundamental representations of
the corresponding gauge groups can arise at the multiple
intersections of such stacks \cite{BDL}. For D4- and D5-branes, to
get $D=4$ {\it chiral} fermions the intersecting branes should sit
at a singular point in the space transverse to the branes, an
orbifold fixed point, for example. In general, intersecting-brane
configurations yield a non-supersymmetric spectrum, so to avoid
the hierarchy problem the string scale associated with such models
must be no more than a few TeV. Gravitational interactions occur
in the bulk ten-dimensional space, and to ensure that the Planck
mass has its observed large value, it is necessary that there are
large dimensions transverse to the branes \cite{ADD}. Thus getting
the correct Planck scale effectively means that only D4- and
D5-brane models are viable, since for D6-branes there is no
dimension transverse to all of the intersecting branes.  However,
a generic feature of these models is that flavour changing neutral
currents are generated by four-fermion operators induced by string
instantons. Although such operators allow the emergence of a
realistic pattern of fermion masses and mixing angles, the severe
experimental limits on flavour changing neutral currents require
that the string scale is rather high , of order $10^4$ TeV
\cite{AO}.  In a non-supersymmetric theory the cancellation of the
closed-string (twisted) Ramond-Ramond (RR) tadpoles does {\it not}
ensure the cancellation of the Neveu-Schwarz-Neveu-Schwarz (NSNS)
tadpoles. There is a resulting instability in the complex
structure moduli \cite{BKLO}. One way to stabilise some of the
(complex structure) moduli is to use an orbifold, rather than a
torus, for the space wrapped by the D-branes.

If the embedding is supersymmetric, then the instabilities
including the gauge hierachy problem are removed. In this case,
one in general has to introduce in addition to D6-branes
orientifold O6-planes, which can be regarded as branes of negative
RR-charge and tension. For a general Calabi-Yau compact space
these orientifold planes wrap special Lagrangian 3-cycles
calibrated with respect to the real part of the holomorphic 3-form
$\Omega_3$ of the Calabi-Yau compact space \footnote{In this case,
the gauge hierarchy problem can be addressed with soft
supersymmetry-breaking terms.}.

This has been studied \cite{Cvetic}, using D6-branes and a
$T^6/(\Z_2\times \Z_2)$ orientifold, but it has so far proved
difficult to get realistic phenomenology consistent with
experimental data from such models. Further progress has been
achieved  using D6-branes and a $\Z_4 \;\cite{BGO},\;\Z_4\times
\Z_2$ \cite{GABRIEL} or $\Z_6$ \cite{Tassilo} orientifold.
Although a semi-realistic three-generation model has been obtained
this way \cite{Tassilo}, it has non-minimal Higgs content, so it
too will have flavour changing neutral currents \cite{DAVIDALEX}
(for recent progress in orientifolds of Gepner models see
\cite{RALPH,HOLLAND}).

An alternative approach in this framework is to start engineering
a grand unified gauge symmetry which subsequently breaks down to
the standard model gauge group \cite{MIRIAM}.  This possibility is
not available in standard type IIB orientifolds, due to the
difficulty in getting adjoint representations to break the GUT
group to the Standard Model \cite{LPT}. A well motivated example
is the flipped $SU(5)\times U(1)_X$ model \cite{BARR,IJDJ}, which
had been extensively studied in the closed string era of the
heterotic compactifications \cite{ANTON,YUAN}. From the
theoretical point of view this motivation was coming from the fact
that its symmetry breaking requires only ${\bf 10}$ and $\bf
{\overline{10}}$ representations at the grand unification scale,
as well as ${\bf 5}$ and ${\bf \bar{5}}$ representations at the
electroweak scale, and these were consistent with the
representations of $SU(5)$ allowed by the unitarity condition with
gauge group at level 1 \cite{PETER,HERBI} \footnote{Thus attempts
to embed conventional grand unified theory (GUT) groups such as
$SU(5)$ or $SO(10)$ in heterotic string required more complicated
compactifications, but none of these has been completely
satisfactory. Constructions with the minimal option to embed just
the standard model gauge group, were plagued with at least extra
unwanted U(1) factors.}. From the phenomenological point of view
flipped $SU(5)\times U(1)_X$ \cite{BARR,IJDJ} has a number of
attractive features in its own right \cite{Dimitri}. For example,
it has a very elegant missing-partner mechanism for suppressing
proton decay via dimension-5 operators \cite{IJDJ}, and is
probably the simplest GUT to survive experimental limits on proton
decay \cite{JDJW}. These considerations motivated the derivation
of a number of flipped $SU(5)$ models from constructions using
fermions on the world sheet \cite{ANTON,YUAN}.  Consistency of the
low energy values of the gauge coupling constants with string
unification at about $10^{18}$ GeV (in the absence of large string
loop threshold corrections) required the existence of extra
matter, besides that of the supersymmetric standard model
\cite{THRE,STUN}.

Non-supersymmetric flipped $SU(5)$ models have been produced in
\cite{PDJ} using D6-branes wrapping toroidal 3-cycles and also
when the wrapping space is the $T^6/\Z_3$ orbifold \footnote{The
$T^6/ \Z_3$ orbifold is not suitable for supersymmetric model
building.}. It is therefore of interest to search for
supersymmetric  flipped $SU(5)$ models from type IIA orientifolds
on $T^6/(\Z_2\times \Z_2)$  with D6-branes intersecting at general
angles.

The wrapping numbers of the various stacks are constrained by the
requirement of RR-tadpole cancellation as well as the
supersymmetry conditions. Tadpole cancellation ensures the absence
of non-abelian anomalies in the emergent low-energy quantum field
theory.  A generalised Green-Schwarz mechanism ensures that 
the gauge bosons associated with all anomalous $U(1)$s acquire
string-scale masses \cite{IRU}, but the gauge bosons of some
non-anomalous $U(1)$s can also acquire string-scale masses
\cite{IMR}; in all such cases the $U(1)$ group survives as a
global symmetry.  Thus we must also ensure the flipped  $U(1)_X$
group remains a gauge symmetry by requiring that its gauge boson
does {\it not} get such a mass.

The material of this Letter is organized as follows. In section 2 we
provide all the necessary formalism for constructing a consistent
string supersymmetric model on the $T^6/(\Z_2\times \Z_2)$
orientifold. This formalism includes the RR tadpole consistency
conditions and the restrictions placed on each stack of D6-branes
for preservation of supersymmetry as well as  the generalised
Green-Schwarz anomaly cancellation mechanism and the requirements
we impose such that the flipped $U(1)_X$ remains a gauge symmetry.

In section 3, for the convenience of the reader,  we first provide
the minimal field-theory content of the flipped $SU(5)$ model and
then we proceed to derive  a string model consistent with the
rules described in section 2. This is a three-generation model,
whose gauge symmetry includes $SU(5)\times U(1)_X$, however with a
non-minimal matter content.

Finally, we use section 4 for our discussions and conclusions.

\section{Search for Supersymmetric Flipped $SU(5)\times U(1)_X $ Brane
Models on a $\mathbf{T^6/(\Z_2\times \Z_2)}$ Orientifold}

   We have several choices at our disposal in
  attempting to build a four-dimensional three-generation
  GUT flipped $SU(5)$ model.  A flipped $SU(5)$ model was
  successfully built in \cite{PDJ} on a $\Z_3$ orientifold but it was
  not supersymmetric.  So, in this paper we
  will focus on the supersymmetric type IIA orientifold on
  $T^6/(\Z_2\times \Z_2)$ with D6-branes intersecting at generic
  angles.  This choice has the feature that $\Z_2$ actions do not
  constrain the ratio of the radii on any 2-torus. Additionally,
  the $T^6/(\Z_2\times \Z_2)$ orbifold
  has only bulk cycles, contrasting the cases of $\Z_4$ and $\Z_6$ orientifolds
  where exceptional cycles also necessarily exist and generally  increase the
  difficulty of satisfying  the Ramond-Ramond tadpole condition.
  However, as we shall see only a limited range of ratio of
  the complex structure moduli is consistent with the
  supersymmetry conditions.

  This $T^6/(\Z_2\times \Z_2)$ structure was first introduced in
  \cite{Cvetic} and further studied in \cite{MIRIAM} \footnote{See also 
\cite{LEIGH}.}, and we will use the
  same notations here.  Consider type IIA theory on the
  $T^6/(\Z_2\times \Z_2)$ orientifold, where the orbifold
  group $\Z_2\times \Z_2$ generators $\theta$, $\omega$ act on the
  complex coordinates $(z_1,z_2,z_3)$ of $T^6=T^2\times T^2\times T^2$ as
\begin{eqnarray}
\theta:(z_1,z_2,z_3)\rightarrow(-z_1,-z_2,z_3) \nonumber \\
\omega:(z_1,z_2,z_3)\rightarrow(z_1,-z_2,-z_3)
\end{eqnarray}
  We implement an orientifold projection $\Omega R$, where
  $\Omega$ is the world-sheet parity, and $R$ acts as
\begin{equation}
R:(z_1,z_2,z_3)\rightarrow(\overline{z}_1,\overline{z}_2,\overline{z}_3)
\end{equation}

  Although the complex structure of the tori is arbitrary under
  the action of $\Z_2\times \Z_2$, it must be assigned consistently
  with the orientifold projection.  Crystallographic action of the
  complex conjugation $R$ restricts consideration to just two shapes.
  We may take either a rectangular toroidal cell or a very specific
  tilted variation.
  Define here a canonical basis of homology cycles ([$a_i$], [$b_i$])
  lying respectively along the $(\hat{x}_i,{\rm i} \hat{y}_i)$ coordinate directions,
  where $i=1,2,3$ labels each of the three $2$-tori.
  Next, consider $K$ different stacks of $N_a$ D6-branes
  wrapping on ([$a_i$], [$b_i$]) with integral coefficients ($n^i_a,m^i_a$),
  where $a=1,2,....K$.
  For the tilted complex structure variants the toroidal cell is
  skewed such that an alternate homology basis is required to
  close cycles spanning the displaced lattice points.
  Specifically, we must consider the cycle
  [$a'_i$]$\equiv$[$a_i$]+$\frac{1}{2}$[$b_i$],
  so that the tilted wrapping is described by
  $n^i_a$[$a'_i$]+$m^i_a$[$b_i$] = $n^i_a$[$a_i$]+$(n^i_a/2+m^i_a)$[$b_i$].
  For convenience, define the effective wrapping number $l^i_a$ as
  $l^i_a\equiv m^i_a$ for rectangular and $l^i_a\equiv 2m^i_a+n^i_a$
  for tilted tori.

With these definitions the homology three-cycles for a stack $a$
of D6-branes and its orientifold image $a'$ are given by
\begin{equation}
[\Pi_a]=\prod_{i=1}^{3}(n^i_a[a_i]+2^{-\beta_i}l^i_a[b_i]),\;\;\;
[\Pi_{a'}]=\prod_{i=1}^{3}(n^i_a[a_i]-2^{-\beta_i}l^i_a[b_i])
\end{equation}
where $\beta_i=0$ if the $i$th torus is not tilted and $\beta_i=1$
if it is tilted.

  There are four kinds of orientifold 6-planes associated with the
actions of $\Omega R$, $\Omega R\theta$, $\Omega R \omega$, and
$\Omega R\theta\omega$. The homology three-cycles which they wrap
are \cite{MIRIAM}
\begin{eqnarray}
&\Omega R : [\Pi_1] = 2^3 [a_1][a_2][a_3],& \Omega R\omega :
[\Pi_2]=-2^{3-\beta_2-\beta_3}[a_1][b_2][b_3] \nonumber \\
&\Omega R\theta\omega : [\Pi_3] =
-2^{3-\beta_1-\beta_3}[b_1][a_2][b_3],& \Omega R\theta :
[\Pi_4]=-2^{3-\beta_1-\beta_2}[b_1][b_2][a_3]
\end{eqnarray}

This represents the fact that $180^{\circ}$ rotation \textit{plus}
conjugate reflection produce `vertical', i.e. $[b_i]$-oriented,
invariant cycles, while the operator $R$ alone preserves certain
cycles along the `horizontal', or $[a_i]$ axis.  Each two-torus
yields always a pair of such cycles, with the exception of the
$[b_i]$-type tilted scenario where only a single invariant
wrapping exists.  This explains then the normal counting of
$8=2^3$ distinct combinations, halved for each application of
tilting in the vertically aligned case.

  The total effect of these four planes should be combined, so we
  define $[\Pi_{O6}]=\sum_i [\Pi_i]$ \cite{MIRIAM}. In addition,
  a set of  new parameters which are convenient
  in the following discussion are introduced \cite{MIRIAM}:
\begin{eqnarray}
& A_a=-n^1_a n^2_a n^3_a,\; B_a=n^1_a l^2_a l^3_a,\; C_a=l^1_a
n^2_a
l^3_a,\; D_a=l^1_a l^2_a n^3_a  \nonumber \\
& \tilde{A_a}=-l^1_a l^2_a l^3_a,\; \tilde{B_a}=l^1_a n^2_a
n^3_a,\; \tilde{C_a}=n^1_a l^2_a n^3_a,\; \tilde{D_a}=n^1_a n^2_a
l^3_a
\label{wrapparameter}
\end{eqnarray}

  With the basic definitions in hand,
  we can continue working on the global constraints of this model.

\subsection{RR-tadpole Consistency Conditions}

The Ramond-Ramond tadpole cancellation requires the total homology
cycle charge of D6-branes and O6-planes to vanish \cite{BERLIN}.
The resulting equation
\begin{equation}
\sum_a N_a[\Pi_a]+\sum_a N_a[\Pi_{a'}]-4[\Pi_{O6}]=0
\end{equation}
can be expressed in terms of the parameters defined in
(\ref{wrapparameter}) as
\begin{equation}
\sum_a N_a A_a=\sum_a N_a B_a=\sum_a N_a C_a=\sum_a N_a D_a = -16
\label{tadpole}
\end{equation}

  It should be stressed that the tadpole condition is independent
  of the selected tilting. However, these coupled constraints are
  generally quite difficult to satisfy. The introduction
  of so called `filler branes' \cite{MIRIAM} which wrap along the O6-planes can
help somewhat. Such branes automatically preserve supersymmetry,
 so that they can be selected with only an eye for independent
 saturation of each RR-tadpole condition. If $N^{(i)}$ branes wrap along the
  $i^{\mathrm{th}}$ O6-plane,  (\ref{tadpole}) is updated to
\begin{eqnarray}
&&-2^k N^{(1)}+\sum_a N_a A_a=-2^k N^{(2)}+\sum_a N_a B_a = \nonumber \\
&&-2^k N^{(3)}+\sum_a N_a C_a=-2^k N^{(4)}+\sum_a N_a D_a = -16
\end{eqnarray}
 Here $k=\beta_1+\beta_2+\beta_3$ is the total
  number of tilted tori.

\subsection{Conditions for Supersymmetric Brane Configurations}

   The condition to preserve $\emph{N}=1$ supersymmetry
  in four dimensions is that the rotation angle of any D-brane with
respect to the orientifold plane is an element of $SU(3)$
\cite{BDL,Cvetic}.  Consider the angles between each brane and the
R-invariant axis of $i^{\mathrm{th}}$ torus $\theta^i_a$, we
require $\theta^1_a +
  \theta^2_a + \theta^3_a = 0$ mod $2\pi$.  This means $\sin(\theta^1_a
+
  \theta^2_a + \theta^3_a)= 0$ and $\cos(\theta^1_a +
  \theta^2_a + \theta^3_a)= 1 > 0 $.  We define
\begin{equation}
\tan\theta^i_a=\frac{2^{-\beta_i}l^i_a R^i_2}{n^i_a R^i_1}
\end{equation}
  where $R^i_2$ and $R^i_1$ are the radii of the $i^{\mathrm{th}}$ torus.
  Then the above supersymmetry conditions can be recast in terms of
the parameters defined in (\ref{wrapparameter}) as follows
\cite{MIRIAM}:
\begin{eqnarray}
x_A\tilde{A_a}+x_B\tilde{B_a}+x_C\tilde{C_a}+x_D\tilde{D_a}=0
\nonumber \\
A_a/x_A + B_a/x_B + C_a/x_C + D_a/x_D <0
\end{eqnarray}
  where $x_A$, $x_B$, $x_C$, $x_D$ are complex structure
  parameters, all of which share the same sign. These parameters
 are given in terms of the complex structure moduli $\chi_i=(R^i_2/R^i_1)$ by
\begin{equation}
x_A=\lambda,\;\;x_B=\lambda2^{\beta_2+\beta_3}/\chi_2\chi_3,
\;\;x_C=\lambda2^{\beta_1+\beta_3}/\chi_1\chi_3,\;\;x_D=\lambda2^{\beta_1+\beta_2}/\chi_1\chi_2
\end{equation}
The positive parameter $\lambda$ was introduced in \cite{MIRIAM} to put
all the variables $A,B,C,D$ on an equal footing. However, among the $x_i$
only three are independent.

\subsection{Intersection Numbers}

The initial $U(N_a)$ gauge group supported by a stack of $N_a$
identical D6-branes is broken down by the $\Z_2\times \Z_2$
symmetry to a subgroup $U(N_a/2)$ \cite{Cvetic}. Chiral matter
particles are formed from  open strings with two ends attaching on
different stacks. By using Grassmann algebra
$[a_i][b_j]=-[b_j][a_i]=\delta_{ij}$ and
$[a_i][a_j]=-[b_j][b_i]=0$ we can calculate the intersection
numbers between stacks $a$ and $b$ and provide  the multiplicity
(${\cal M}$) of the corresponding bi-fundamental representation:
\begin{equation}
{\cal M}(\frac{N_a}{2},
\frac{\overline{N_b}}{2})=I_{ab}=[\Pi_a][\Pi_b]=2^{-k}\prod_{i=1}^3(n^i_a
l^i_b - n^i_b l^i_a)
\end{equation}
Likewise, stack $a$ paired with the orientifold image $b'$ of $b$
yields
\begin{equation}
{\cal M}(\frac{N_a}{2},
\frac{N_b}{2})=I_{ab'}=[\Pi_a][\Pi_{b'}]=-2^{-k}\prod_{i=1}^3(n^i_a
l^i_b + n^i_b l^i_a)
\end{equation}

Strings stretching between a brane in stack $a$ and its mirror
image $a'$ yield chiral matter in the antisymmetric and symmetric
representations of the group $U(N_a/2)$ with multiplicities
\begin{equation}
{\cal M}(({\rm A}_a)_L)=\frac{1}{2}I_{aO6},\;\; {\cal M}(({\rm
A}_a+{\rm S}_a)_L)=\frac{1}{2}(I_{aa'}-\frac{1}{2}I_{aO6})
\end{equation}
so that the net total of antisymmetric and symmetric
representations are given by
\begin{eqnarray}
{\cal M}({\rm Anti}_a)=\frac{1}{2}(I_{aa'}+\frac{1}{2}I_{aO6}) =
-2^{1-k}[(2A_a-1)\tilde{A_a}-\tilde{B_a}-\tilde{C_a}-\tilde{D_a}]
\nonumber \\ {\cal M}({\rm
Sym}_a)=\frac{1}{2}(I_{aa'}-\frac{1}{2}I_{aO6}) =
-2^{1-k}[(2A_a+1)\tilde{A_a}+\tilde{B_a}+\tilde{C_a}+\tilde{D_a}]
\label{netmult}
\end{eqnarray}
where
\begin{equation}
I_{aa'}=[\Pi_a][\Pi_{a'}]=-2^{3-k}\prod_{i=1}^3 \,n^i_a l^i_a
\end{equation}
\begin{equation}
I_{aO6}=[\Pi_a][\Pi_{O6}]=2^{3-k}(\tilde{A_a}+\tilde{B_a}+\tilde{C_a}+\tilde{D_a})
\end{equation}

This distinction is critical, as we require independent use of the
paired multiplets such as $(\mathbf{10},\mathbf{\overline{10}})$
which are masked in expression (\ref{netmult}).  In what follows
we consider the case $k=0$.

\subsection{Generalized Green-Schwarz Mechanism}

   Although the total non-Abelian
  anomaly in intersecting brane world models cancels automatically when the
  RR-tadpole conditions are satisfied, there may be additional mixed
  anomalies present.  For instance, the mixed gravitational
  anomalies which generate massive fields are not trivially zero \cite{Cvetic}.
  These anomalies are cancelled by a generalized
  Green-Schwarz (G-S) mechanism which involves untwisted
  Ramond-Ramond forms.  The couplings of the four untwisted
  Ramond-Ramond forms $B^i_2$
  to the $U(1)$ field strength $F_a$ of each stack $a$ are
\begin{eqnarray}
 N_a l^1_a n^2_a n^3_a \int_{M4}B^1_2\wedge \textrm{tr}F_a,  \;\;
 N_a n^1_a l^2_a n^3_a \int_{M4}B^2_2\wedge \textrm{tr}F_a
  \nonumber \\
 N_a n^1_a n^2_a l^3_a \int_{M4}B^3_2\wedge \textrm{tr}F_a,  \;\;
-N_a l^1_a l^2_a l^3_a \int_{M4}B^4_2\wedge \textrm{tr}F_a
\end{eqnarray}

  These couplings determine the linear combinations of $U(1)$
  gauge bosons that acquire string scale masses via the G-S
  mechanism.  In flipped $SU(5)\times U(1)_X$,  the symmetry
 $U(1)_X$ must remain a gauge symmetry  so that it
may remix to help generate the standard model hypercharge after
the breaking of $SU(5)$.  Therefore, we must ensure that the gauge
boson of the flipped $U(1)_X$ group does not receive such a mass.
The $U(1)_X$ is a linear combination (to be identified in section
3.2) of the $U(1)$s from each stack :
\begin{equation}
U(1)_X=\sum_a C_a U(1)_a
\end{equation}
The
corresponding field strength must be orthogonal to those that
acquire G-S mass.  Thus we demand :
\begin{eqnarray}
\sum_a C_a N_a \tilde{B_a} =0, \;\; \sum_a C_a N_a \tilde{C_a} =0
  \nonumber \\
\sum_a C_a N_a \tilde{D_a} =0, \;\; \sum_a C_a N_a \tilde{A_a} =0
\label{GSeq}
\end{eqnarray}

\section{Flipped $SU(5)\times U(1)_X$ Model Building}

In the previous section we have outlined all the necessary
machinery for constructing an intersecting-brane model on the
$T^6/(\Z_2\times \Z_2)$ orientifold.  Our goal now is to realize a
supersymmetric $SU(5)\times U(1)_X$ gauge theory with three
generations and a complete GUT and electroweak Higgs sector in the
four-dimensional spacetime. We also try to avoid as much extra
matter as possible.
\subsection{Basic Flipped $SU(5)$ Phenomenology}

In a flipped $SU(5)\times U(1)_X$ \cite{BARR,IJDJ} unified model,
the electric charge generator $Q$ is only partially embedded in
$SU(5)$, {\it i.e.}, $Q = T_3 - \frac{1}{5}Y' +
\frac{2}{5}\tilde{Y}$, where $Y'$ is the $U(1)$ internal $SU(5)$
and $\tilde{Y}$ is the external $U(1)_X$ factor.  Essentially,
this means that the photon is \lq shared\rq \ between $SU(5)$ and
$U(1)_X$. The Standard Model (SM) plus right handed neutrino
states reside within the representations $\bar{\bf{5}}$,
$\bf{10}$, and $\bf{1}$ of $SU(5)$, which are collectively
equivalent to a spinor $\bf{16}$ of $SO(10)$.  The quark and
lepton assignments are flipped by $u^c_L$ $\leftrightarrow$
$d^c_L$ and $\nu^c_L$ $\leftrightarrow$ $e^c_L$ relative to a
conventional $SU(5)$ GUT embedding:
\begin{equation}
\bar{f}_{\bf{\bar{5},-\frac{3}{2}}}= \left( \begin{array}{c}
              u^c_1 \\ u^c_2 \\ u^c_3 \\ e \\ \nu_e
                    \end{array} \right) _L ; \;\;\;
F_{\bf{10,\frac{1}{2}}}= \left( \left( \begin{array}{c}
              u \\ d \end{array} \right) _L  d^c_L \;\; \nu^c_L
\right)
              ; \;\;\;
l_{\bf{1,\frac{5}{2}}}=e^c_L
\end{equation}
In particular this results in  the $\bf{10}$ containing a neutral component with the
quantum numbers of $\nu^c_L$.  We can break spontaneously the GUT
symmetry by using a $\bf{10}$ and $\overline{\bf{10}}$ of
superheavy Higgs where the neutral components provide a large
vacuum expectation value, $\left\langle \nu^c_H \right\rangle$=
$\left\langle \bar{\nu}^c_H \right\rangle$,
\begin{equation}
H_{\bf{10,\frac{1}{2}}}=\left\{Q_H,\;d^c_H,\;\nu^c_H \right\}; \;\;\;
\bar{H}_{\bf{\overline{10},-\frac{1}{2}}}=\left\{Q_{\bar{H}},\;d^c_{\bar{H}},\;\nu^c_{\bar{H}}
\right\}.
\end{equation}
The electroweak spontaneous breaking is generated by the Higgs
doublets $H_2 $ and $ \bar{H}_{\bar{2}} $
\begin{equation}
h_{\bf{5,-1}}=\left\{ H_2,H_3 \right\}; \;\;\;
\bar{h}_{\bf{\bar{5},1}}=\left\{
\bar{H}_{\bar{2}},\bar{H}_{\bar{3}} \right\}
\end{equation}
Flipped $SU(5)$ model building has two very nice features which
are generally not found in typical unified models: (i) a natural
solution to the doublet ($H_2$)-triplet($H_3$) splitting problem
of the electroweak Higgs pentaplets $h,\bar{h}$ through the
trilinear coupling of the Higgs fields: $H_{\bf{10}} \cdot
H_{\bf{10}} \cdot h_{\bf{5}} \rightarrow \left\langle \nu^c_H
\right\rangle d^c_H H_3$, and (ii) an automatic see-saw mechanism
that provide heavy right-handed neutrino mass through the coupling
to singlet fields $\phi$, $F_{\bf{10}} \cdot {\bar
H}_{\overline{\bf{10}}} \cdot \phi \rightarrow \left\langle
\nu^c_{\bar{H}}\right\rangle \nu^c \phi$.

The generic superpotential $W$ for a flipped $SU(5)$ model will be of the form :
\begin{equation}
\lambda_1 FFh+\lambda_2 F\bar{f}\bar{h}+\lambda_3 \bar{f}l^c h+ \lambda_4 F\bar{H}\phi +\lambda_5 HHh+\lambda_6 \bar{H}\bar{H}\bar{h}+ \cdots\in W
\end{equation}
the first three terms provide masses for the quarks and leptons,
the fourth is responsible for the heavy right-handed neutrino mass
and the last two terms are responsible for the doublet-triplet
splitting mechanism \cite{IJDJ}.

\subsection{Model Building}

We first consider a stack with ten D6-branes to form the desired
$U(5)$ group, and then determine additional stacks of two branes
which provide $U(1)$ group factors and are compatible with the
supersymmetry conditions of the 10-brane stack.  To have enough
but not too many copies of the antisymmetric and symmetric
representation in the first stack $a$ to satisfy the tadpole
conditions, it is reasonable to consider the case of no tilted
tori ($k=0$) and we choose a set of proper wrapping numbers to
make ${\cal M}(({\rm A}_a)_L)=4$ and ${\cal M}(({\rm A}_a+{\rm
S}_a)_L)=-2$.  Under this setting, one wrapping number is zero and
it makes two of the RR-tadpole parameters $A$, $B$, $C$, $D$ zero
with the remaining two negative, which forces the structure
parameters $x_A$, $x_B$, $x_C$, $x_D$ to be all positive by the
SUSY conditions.  Then the rest of the 2-brane stacks are chosen
in accordance with our requirements.

Because of the combined constraints from RR-tadpole and  SUSY
conditions, it is harder to get negative values than to get
positive values or zero for $I_{ab}$ and $I_{ab'}$ to generate the
required bi-fundamental representations.  Generally when a
negative number is needed, the absolute value cannot be large
enough to alone provide three generations of chiral matter.  This
suggests the consideration of multiple two-brane stacks to share
the burden of this task.

Next we turn to the question of the number of stacks we need.
Generally speaking a case with three stacks is enough to provide
all the required matter to construct a normal $SU(5)$ GUT model.
However, as we mentioned we have to ensure that the $U(1)_X$
remains a gauge  symmetry after the application of the G-S
mechanism.  It is clear that at least two more stacks are needed
if all the couplings to the four RR forms are present.

The pentaplet $\bar{f} $ which contains Standard Model fermions is different
from the Higgs pentaplet $\bar{h}$ resulting from the
`flipped' nature of the model  as we saw in section 3.1.
For example, if we take $U(1)_X$ for
  ${\bf (10,1)}$ in both SM and Higgs spectrum as $1/2$, then it is $-3/2$ for
  ${\bf (\bar{5},1)}$ in SM, $5/2$ for ${\bf (1,1)}$ in SM, $-1/2$ for
  ${\bf (\overline{10},1)}$ in Higgs, 1 or -1 for ${\bf (\bar{5},1)}$ and
  ${\bf (5,1)}$ in Higgs, and 0 for ${\bf (1,1)}$ in Higgs.  These constrain
some coefficients of $U(1)$s from the stacks involving the SM and
Higgs spectra, and may require more stacks in addition to the five
mentioned above for obtaining the correct $U(1)_X$ charge for all
the matter and Higgs representations. In this paper we present an
example with seven stacks.

However, with seven stacks it was still difficult to find chiral
bi-fundamental  representations to be identified with the
electroweak Higgs pentaplets, $h,\bar{h}$ and at the same time for
the $U(1)_X$ group to remain a gauge symmetry.
 This directed us towards the most natural choice of identifying our Higgs
pentaplets as well as some matter representations from
intersections which provide non-chiral matter.  After all, the
Higgs $\mathbf{5}$ and the $\mathbf{\bar{5}}$ construct the
vector-like $\mathbf{10}$ representation of $SO(10)$. A zero
intersection number between two branes implies that the branes are
parallel on at least one torus.  At such kind of intersection
additional non-chiral (vector-like) multiplet pairs from $ab+ba$,
$ab'+b'a$, and $aa'+a'a$ can arise
\cite{PAUL}\footnote{Representations $({\rm Anti}_a+\overline{\rm
Anti}_a)$ occur at intersection of $a$ with $a^{\prime}$ if they
are parallel on at least one torus.}.  The multiplicity of these
non-chiral multiplet pairs is given by the remainder of the
intersection product, neglecting the null sector.  For example, if
$(n^1_a l^1_b - n^1_b l^1_a)=0$ in $
I_{ab}=[\Pi_a][\Pi_b]=2^{-k}\prod_{i=1}^3(n^i_a l^i_b - n^i_b
l^i_a) $,
\begin{equation}
{\cal M}\left[\left(\frac{N_a}{2},\frac{\overline{N_b}}{2}\right)
+\left(\frac{\overline{N_a}}{2},\frac{N_b}{2}\right)\right]
=\prod_{i=2}^3(n^i_a l^i_b - n^i_b l^i_a)
\end{equation}
This is useful since we can fill the spectrum with this matter
without affecting the required global conditions because the total
effect of the pairs is zero. For instance in our model, besides
the $(ae^{\prime})$ intersection which provides a vector-like pair
of Higgs pentaplets, the intersection $(ef^{\prime})$ delivers the
fermion (singlet under the $SU(5)$ group) $l_{\bf 1,\frac{5}{2}}$
particles.

In table 1 we present a consistent model compatible with the
constraints we described. Note that this is a (7+1)-stack model,
with one stack of two filler branes wrapped along the first
orientifold plane and two sets of parallel branes; the latter
provide several non-chiral pairs. The gauge symmetry associated
with the two filler branes is $Usp(2)\cong SU(2)$.

\paragraph{The Result}
The gauge symmetry of the (7+1)-stack model in table 1 is
$U(5)\times U(1)^6\times Usp(2)$, and the structure parameters of
the wrapping space are
\begin{equation}
x_A=1, \;\; x_B=2, \;\; x_C=8, \;\; x_D=1
\end{equation}
which means
\begin{equation}
\frac{R^1_2}{R^1_1}=\frac{1}{2}\,, \;\; \frac{R^2_2}{R^2_1}=2,
\;\; \frac{R^3_2}{R^3_1}=\frac{1}{4}
\end{equation}

  The intersection numbers are listed in table 2, and the resulting spectrum
  in table 3.  We have a complete Standard Model sector plus right handed neutrinos in three copies,  a
  complete Higgs spectrum, and in addition extra $exotic$ matter
  which includes two ${\bf (\overline{15},1)}$.

The $U(1)_X$ is
\begin{equation}
U(1)_X=\frac{1}{12}(3U(1)_a-20U(1)_b+45U(1)_d-15U(1)_e-15U(1)_f-20U(1)_g)
\end{equation}
while the other two anomaly-free and massless combinations $U(1)_Y$ and
$U(1)_Z$ are
\begin{eqnarray}
U(1)_Y & = & U(1)_b+U(1)_c-6U(1)_d+3U(1)_e+3U(1)_f+2U(1)_g
\nonumber \\
U(1)_Z & = & U(1)_b-U(1)_c+U(1)_e-U(1)_f
\end{eqnarray}
These two  gauge symmetries  can be spontaneously broken
by assigning vacuum expectation values to singlets from the intersection
$(bg)$. Thus, the final gauge symmetry is $SU(5)\times U(1)_X \times Usp(2)$.

The remaining four global $U(1)$s from the Green-Schwarz mechanism
are given respectively by
\begin{eqnarray}
 U(1)_1 & = & -10U(1)_a+2U(1)_b+2U(1)_c-2U(1)_d-8U(1)_g   \nonumber\\
 U(1)_2 & = & -2U(1)_b-2U(1)_c+2U(1)_g   \nonumber\\
 U(1)_3 & = & 6U(1)_b+6U(1)_c+4U(1)_d+2U(1)_e+2U(1)_f  \nonumber\\
 U(1)_4 & = & 20U(1)_a+6U(1)_b+6U(1)_c-2U(1)_e-2U(1)_f.
\end{eqnarray}

\begin{table}[h]
\begin{tabular}{|c|c||c|c|c||c|c|c|c||c|c|c|c|}
\hline

stack& $N_a$ & ($n_1$, $l_1$) & ($n_2$, $l_2$) & ($n_3$, $l_3$) & $A$ &
$B$ & $C$ & $D$ & $\tilde{A}$ & $\tilde{B}$ & $\tilde{C}$ &
$\tilde{D}$  \\ \hline \hline

$a$ & $N=10$ & ( 0,-1) &(-1,-1) & (-1,-2) & 0 & 0 & -2 & -1 & 2
& -1 & 0 & 0 \\ \hline

 $b$ & $N=2$ & (-1,-1) & (-1, 1) & ( 1, 3) & -1 & -3 & 3 & -1 & 3 &
1 & -1 & 3  \\ \hline

 $c$ & $N=2$ & (-1,-1) & (-1, 1) & ( 1, 3)& -1 & -3 & 3 & -1 & 3 &
1 & -1 & 3  \\ \hline

 $d$ & $N=2$ & (-1, 1) &( 1, 0) & (-1,-2) & -1 & 0 & -2 & 0 & 0 & -1
& 0 & 2  \\ \hline

 $e$ & $N=2$ & (-1, 1) & ( 1,-1) & ( 0,-1) & 0 & -1 & -1 & 0 & -1 &
0 & 0 & 1  \\ \hline

 $f$ & $N=2$ & (-1, 1) & ( 1,-1) & ( 0,-1) & 0 & -1 & -1 & 0 & -1 &
0 & 0 & 1  \\ \hline

 $g$ & $N=2$ & ( 1,-1) & (-4,-1) & (-1, 0) & -4 & 0 & 0 & -1 & 0 &
-4 & 1 & 0  \\ \hline

filler & $N^{(1)}=2$ & ( 1, 0) & ( 1, 0) & ( 1, 0) & -1 & 0 & 0 & 0 & 0 &
0 & 0 & 0
\\ \hline
\end{tabular}
\caption{Wrapping numbers and their consistent parameters.}
\end{table}

\begin{table}[h]
\begin{center}
\footnotesize
\begin{tabular}{|c|c||c|c||c|c|c|c|c|c|c|c|c|c|c|c||c|}
\hline

stk & $N$ & A & S & $b$ & $b'$ & $c$ & $c'$ & $d$ & $d'$ & $e$ &
$e'$ & $f$ & $f'$ & $g$ & $g'$ & f1  \\ \hline \hline

$a$ & 10 & 2 & -2 & -2 & 0(5) & -2 & 0(5) & 0(1) & 4 & -2 & 0(1) &
-2 & 0(1) & 6 & 10 & 2  \\ \hline

$b$ & 2 & 24 & 0 & - & - & 0(0) & 24 & 2 & 0(5) & 0(2) & 0(2) &
0(2) & 0(2) & 30 & 0(9) & 3  \\ \hline

$c$ & 2 & 24 & 0 & - & - & - & - & 2 & 0(5) & 0(2) & 0(2) & 0(2) &
0(2) & 30 & 0(9) & 3  \\ \hline

$d$ & 2 & 2 & -2 & - & - & - & - & - & - & 0(1) & -2 & 0(1) & -2 &
0(2) & 4 & 0  \\ \hline

$e$ & 2 & 0 & 0 & - & - & - & - & - & - & - & - & 0(0) & 0(4) &
0(5) & -6 & -1  \\ \hline

$f$ & 2 & 0 & 0 & - & - & - & - & - & - & - & - & - & - & 0(5) &
-6 & -1  \\ \hline

$g$ & 2 & -6 & 6 & - & - & - & - & - & - & - & - & - & - & - & - &
0  \\ \hline
\end{tabular}
\caption{List of intersection numbers.  The number in parenthesis
indicates the multiplicity of non-chiral pairs.}
\end{center}
\end{table}

\begin{table}[h]
\begin{center}
\footnotesize
\begin{tabular}{|c||@{}c@{}||@{}c@{}|@{}c@{}|@{}c@{}|@{}c@{}|@{}c@{}|@{}c@{}|@{}c@{}||@{}c@{}||@{}c@{}|
@{}c@{}|@{}c@{}|@{}c@{}||@{}c@{}|@{}c@{}|} \hline

 Rep. & Multi. &$U(1)_a$&$U(1)_b$&$U(1)_c$& $U(1)_d$
& $U(1)_e$ & $U(1)_f$ & $U(1)_g$ & $12U(1)_X$ & $U(1)_1$ & $U(1)_2$ &
$U(1)_3$ & $U(1)_4$ & $U(1)_Y$ & $U(1)_Z$ \\
\hline \hline

$(10,1)$ & 3 & 2 & 0 & 0 & 0 & 0 & 0 & 0 & 6 & -20 & 0 & 0 & 40 & 0 & 0
\\

$(\bar{5}_a ,1_e)$ & 2 & -1 & 0 & 0 & 0 & 1 & 0 & 0 & -18 & 10 & 0 & 2
& -22 & 3 & 1  \\

$(\bar{5}_a ,1_f)$ & 1 & -1 & 0 & 0 & 0 & 0 & 1 & 0 & -18 & 10 & 0 & 2
& -22 & 3 & -1  \\
\hline

$(\bar{1}_e ,\bar{1}_f)^\star$ & 3 & 0 & 0 & 0 & 0 & -1 & -1 & 0 &
30 & 0 & 0 & -4 & 4 & -6 & 0
\\ \hline \hline

$(10,1)$ & 1 & 2 & 0 & 0 & 0 & 0 & 0 & 0 & 6 & -20 & 0 & 0 & 40 & 0 & 0
\\

$(\overline{10},1)$ & 1 & -2 & 0 & 0 & 0 & 0 & 0 & 0 & -6 & 20 & 0 & 0
& -40 & 0 & 0  \\
\hline

$(5_a,1_e)^\star$ & 1 & 1 & 0 & 0 & 0 & 1 & 0 & 0 & -12 & -10 & 0 & 2 &
18 & 3 & 1   \\

$(\bar{5}_a ,\bar{1}_e)^\star$ & 1 & -1 & 0 & 0 & 0 & -1 & 0 & 0 &
12 & 10 & 0 & -2 & -18 & -3 & -1
\\ \hline

$(1_b ,\bar{1}_g)$ & 4 & 0 & 1 & 0 & 0 & 0 & 0 & -1 & 0 & 10 & -4 & 6 &
6 & -1 & 1   \\
\hline \hline

$(\overline{15},1)$ & 2 & -2 & 0 & 0 & 0 & 0 & 0 & 0 & -6 & 20 & 0 & 0
& -40 & 0 & 0   \\

$(\overline{10},1)$ & 1 & -2 & 0 & 0 & 0 & 0 & 0 & 0 & -6 & 20 & 0 & 0
& -40 & 0 & 0   \\
\hline

$(\bar{5}_a ,1_c)$ & 2 & -1 & 0 & 1 & 0 & 0 & 0 & 0 & -3 & 12 & -2 & 6
& -14 & 1 & -1   \\

$(5_a ,1_d)$ & 4 & 1 & 0 & 0 & 1 & 0 & 0 & 0 & 48 & -12 & 0 & 4 & 20 &
-6 & 0    \\

$(\bar{5}_a ,1_b)$ & 2 & -1 & 1 & 0 & 0 & 0 & 0 & 0 & -23 & 12 & -2 & 6
& -14 & 1 & 1   \\

$(\bar{5}_a ,1_f)$ & 1 & -1 & 0 & 0 & 0 & 0 & 1 & 0 & -18 & 10 & 0 & 2
& -22 & 3 & -1    \\

$(5_a ,\bar{1}_g)$ & 6 & 1 & 0 & 0 & 0 & 0 & 0 & -1 & 23 & -2 & -2 & 0
& 20 & -2 & 0    \\

$(5_a ,1_g)$ & 10 & 1 & 0 & 0 & 0 & 0 & 0 & 1 & -17 & -18 & 2 & 0
& 20 & 2 & 0   \\ \hline

$(1_b ,1_c)$ & 24 & 0 & 1 & 1 & 0 & 0 & 0 & 0 & -20 & 4 & -4 & 12 & 12
& 2 & 0   \\

$(1_b ,\bar{1}_d)$ & 2 & 0 & 1 & 0 & -1 & 0 & 0 & 0 & -65 & 4 & -2 & 2
& 6 & 7 & 1   \\

$(1_b ,\bar{1}_g)$ & 26 & 0 & 1 & 0 & 0 & 0 & 0 & -1 & 0 & 10 & -4 & 6
& 6 & -1 & 1   \\

$(1_c ,\bar{1}_d)$ & 2 & 0 & 0 & 1 & -1 & 0 & 0 & 0 & -45 & 4 & -2 & 2
& 6 & 7 & -1   \\

$(1_c ,\bar{1}_g)$ & 30 & 0 & 0 & 1 & 0 & 0 & 0 & -1 & 20 & 10 & -4 & 6
& 6 & -1 & -1   \\

$(\bar{1}_d ,\bar{1}_e)$ & 2 & 0 & 0 & 0 & -1 & -1 & 0 & 0 & -30 &
2 & 0 & -6 & 2 & 3 & -1  \\
$(\bar{1}_d ,\bar{1}_f)$ & 2 & 0 & 0 & 0 & -1 & 0 & -1 & 0 & -30 &
2 & 0 & -6 & 2 & 3 & 1  \\
$(1_d ,1_g)$ & 4 & 0 & 0 & 0 & 1 & 0 & 0 & 1 & 25 & -10 & 2 & 4 & 0 &
-4 & 0   \\

$(\bar{1}_e ,\bar{1}_g)$ & 6 & 0 & 0 & 0 & 0 & -1 & 0 & -1 & 35 &
8 & -2 & -2 & 2 & -5 & -1  \\
$(\bar{1}_f ,\bar{1}_g)$ & 6 & 0 & 0 & 0 & 0 & 0 & -1 & -1 & 35 &
8 & -2 & -2 & 2 & -5 & 1  \\
$(\bar{1},\bar{1})$ & 2 & 0 & 0 & 0 & -2 & 0 & 0 & 0 & -90 &
4 & 0 & -8 & 0 & 12 & 0   \\

$(1,1)$ & 6 & 0 & 0 & 0 & 0 & 0 & 0 & 2 & -40 & -16 & 4 & 0 & 0 &
4 & 0  \\ \hline

$(1_e ,1_f)^\star$ & 4 & 0 & 0 & 0 &0 & 1 & 1 & 0 & -30 &
0 & 0 & 4 & -4 & 6 & 0   \\

$(\bar{1}_e ,\bar{1}_f)^\star$ & 1 & 0 & 0 & 0 & 0 & -1 & -1 & 0 &
30 & 0 & 0 & -4 & 4 & -6 & 0  \\
\hline

\multicolumn{16}{|c|}{Additional non-chiral Matter}\\ \hline
\multicolumn{16}{|c|}{Usp(2) Matter}\\ \hline\hline
\end{tabular}
\caption{The spectrum of $U(5)\times U(1)^6\times Usp(2)$, or
$SU(5)\times U(1)_X\times U(1)_Y\times U(1)_Z\times Usp(2)$, with
the four global $U(1)$s from the Green-Schwarz mechanism.  The
$\star'd$ representations stem from vector-like non-chiral pairs.}
\end{center}
\end{table}

\clearpage

\section{Conclusions}

In this Letter we have constructed a particular $N=1$
supersymmetric three-family model whose gauge symmetry includes
$SU(5)\times U(1)_X$, from type IIA orientifolds on $
T^6/(\Z_2\times \Z_2)$ with D6-branes intersecting at general
angles. The spectrum contains a complete grand unified theory and
electroweak Higgs sector. In addition, it contains extra exotic
matter both in bi-fundamental and vector-like representations as
well as two copies of matter in the symmetric representation of
$SU(5)$. Chiral matter charged under both the $SU(5)\times U(1)_X$
and $USp(2)$ gauge symmetries is also present, as is evident from
Table 2. Furthermore, three adjoint $(N=1)$ chiral multiplets are
provided from the $aa$ sector \cite{Cvetic}. We also note that 
the low energy spectrum of the model we constructed is free from any 
$SU(2)$ global anomalies since the number of the corresponding 
fermion doublets is even \cite{Witten}. Nevertheless, although the massless 
spectrum  
is free from such global anomalies it does not satisfy all the additional
 constraints 
arising from the K-theory interpretation of D-branes \cite{SCM, MC}. 
This issue will be investigated in a future publication.

The global symmetries, that arise after the G-S anomaly
cancellation mechanism, forbid some of the Yukawa couplings
required for mass generation, for instance terms like $FFh$.
However, by the same token the term $HHh$ is also forbidden. We
note that such a term is essential for the doublet-triplet
splitting solution mechanism in flipped $SU(5)$.  Nevertheless, it
should not escape our notice that while these global $U(1)$
symmetries are exact to all orders in perturbation theory, they
can be broken explicitly by non-perturbative instanton
effects \cite{REVIEWS1,kachru}. Thus, providing us with the possibility of
recovering the appropriate superpotential couplings. Another
interesting approach toward generating these absent Yukawa
couplings may entail the introduction of type IIB flux
compactifications \cite{cjan05}. This exceeds the scope of our
current Letter, but shall be further investigated in an upcoming
publication \cite{nkwcm}.

\section*{Acknowledgements}
We would like to thank R. Blumenhagen for useful correspondence. 
We also thank M.  Cveti\v{c} and G. Shiu for prompting our 
attention to the constraints arising from K-theory.
The work of G.V.K. and D.V.N. is supported by DOE grant
DE-FG03-95-Er-40917.

\end{document}